\documentclass[useAMS,usenatbib]{mn2e}
\usepackage{graphics}
\usepackage{epsfig}
\input psfig.sty

\begin {document}

\title[How galaxies lose their angular momentum]
 {How galaxies lose their angular momentum}
\author[E. D'Onghia, A. Burkert, G. Murante \& S. Khochfar]
{Elena~D'Onghia$^{1,4}$\thanks{donghia@usm.uni-muenchen.de}, Andreas~Burkert$^{1}$, Giuseppe~Murante$^{2}$ \& Sadegh~Khochfar$^{3}$\\
 $^1$ University Observatory Munich, Scheinerstr. 1, 81679 Munich, Germany \\
 $^2$ INAF Osservatorio Astronomico di Torino, Strada dell'Osservatorio 20 I-10025 
 Pino Torinese, Italy\\
 $^3$ Department of Physics, Denys Wilkinson Building, Keble Road, Oxford OX1 3RH, 
 United Kingdom \\
 $^4$ Max-Planck-Institut f\"ur  Extraterrestrische Physik, Giessenbachstr., 
 85748 Garching, Germany\\ }
	     
\date{submitted to MNRAS}
	     
\pagerange{\pageref{firstpage}--\pageref{lastpage}}\pubyear{2006}
\maketitle	     

\label{firstpage}
	     
\begin{abstract}
The processes are investigated by which gas loses its angular
momentum during the protogalactic collapse
phase, leading to disk galaxies that are too compact with respect to the
observations. High-resolution N-body/SPH simulations in a
cosmological context are presented
including cold gas and dark matter. A halo with quiet merging
activity since redshift z$\sim$3.8 and with 
a high spin parameter is analysed that should be an ideal candidate
for the formation of an extended galactic disk.
We show that the gas and the dark matter have similar specific angular
momenta until a merger event occurs at z$\sim$2 with a mass ratio of 5:1.
All the gas involved in the merger loses a substantial fraction of its specific
angular momentum due to tidal torques and falls quickly into the centre. 
Dynamical friction 
plays a minor role, in contrast to previous claims. In fact, after this
event a new extended disk begins to form from gas
that was not involved in the 5:1 merger event and that falls in subsequently.
We argue that the angular momentum problem of disk galaxy formation is a
merger problem: in cold dark matter cosmology substantial mergers with
mass ratios of 1:1 to 6:1 are expected to occur in almost all galaxies.
We suggest that energetic feedback processes could in principle
solve this problem, however only if the heating occurs at the time or shortly
before the last substantial merger event. Good candidates for such a coordinated
feedback would be a merger-triggered star burst or central black hole heating. 
If a large fraction of the
low angular momentum gas would be ejected, late-type
galaxies could form with a dominant extended disk component, resulting from 
late infall, a small bulge-to-disk ratio and a low baryon fraction, in 
agreement with observations.

\end{abstract}


\setcounter{footnote}{1}

\section{Introduction}
\label{sec:introduction}
An important goal of cosmology is to understand 
how galaxies form. Currently,
the most popular scenario for structure formation in the Universe is based on
the inflationary cold dark matter (CDM) theory (Blumenthal et
al. 1984), according to which cosmic structures arise from
small Gaussian density fluctuations composed of non-relativistic
collisionless particles. The dark matter aggregates into larger and larger
clumps as gravity amplifies the weak density perturbations, produced
at early times in the universe.  Gas cools and condenses within these dark halos,
eventually forming the galaxies we see today (White $\&$ Rees 1978).
This hierarchical picture of structure formation is an elegant and well
defined theory that naturally explains the growth of large-scale structures
from density fluctuations as small as those detected in the Cosmic Microwave
Background to present-day galaxies. 
The CDM model however also has problems that might at the end lead to a deeper understanding
of the nature and origin of cold dark matter and its interaction with baryonic matter.
One of the most interesting puzzles at the moment is the so called cosmological 
angular momentum problem (for a review see e.g. Burkert \& D'Onghia 2004, Primack 2005).

In spiral galaxies, almost all the stars and the gas move on circular orbits. The structure
of these galaxies is therefore governed by angular momentum.
Where does this angular momentum come from? According to the standard picture,
the angular momentum of a dark matter fluctuation grows by tidal torques from
neighboring structures (Peebles 1969, White 1984), until the protogalaxy
decouples from the Hubble flow and collapses.
Fall $\&$ Efstathiou (1980) argued that gas and dark matter should initially
have been well mixed. In this case, the specific angular momentum distribution
of the gas is initially equal to that of the dark halo.
During the protogalactic collapse phase the gas component dissipates its
kinetic energy, decouples from the collisionless and violently relaxing dark
matter component and settles into a flattened, fast rotating disk that
subsequently turns into stars.
If the specific angular momentum were preserved during the infall into the
equatorial plane, the scale length of the galactic disk would be directly
related to the specific angular momentum of the halo. Fall $\&$ Efstathiou
(1980) and later on Mo, Mao $\&$ White (1998) indeed showed that
in this case the expected
disk scale lengths are in good agreement with the observations (e.g.
 Courteau 1997).

In contrast to these analytical estimates, 
numerical simulations produce disks that are
too small and much more centrally concentrated than observed.
This is the so called angular momentum catastrophy of the gas (Navarro $\&$ Benz 1991; Navarro $\&$
Steinmetz 1997). Navarro $\&$ Steinmetz (2000) suggested that the loss of
angular momentum is a result of dense gaseous clumps falling into the inner regions
of dark halos and transfering large amounts of their angular momentum to the dark
matter component by dynamical friction. However this process has
not been quantified and investigated in details.

Another, probably related problem, encountered by numerical simulations
of galaxy formation is that without energetic heating star formation is already very
efficient in low mass structures at high redshifts (Navarro, Frenk $\&$ White 1995,
Steinmetz $\&$ M\"uller 1994). The dense, compact stellar systems, formed
in that way, later are collected in the innermost regions of larger galaxies,
forming compact bulge-dominated systems that do not resemble real late-type spiral
galaxies.  Subsequent work tried to solve some of these shortcomings by employing
energetic feedback, leading to more promising results
(Sommer-Larsen, G\"otz $\&$ Portinari 2003; Abadi et al. 2003; Governato
et al. 2004; Robertson et al. 2004). Here, the
idea was that heating would reduce early star formation and, at the same time,
decouple gas from the dark haloes in an early
phase,  thus also reducing the effect of dynamical friction and angular momentum loss
on the gas component when the dark matter substructures merge. Lateron,
gas cools and falls back into the equatorial plane in a monolithic-like collapse.

Although some cases have now been reported to form realistic disk galaxies
in this way, most disks
are still denser and contain more massive bulges than observed in late-type
galaxies, indicating that the specific angular momentum problem is not solved.
It also has been suggested that only halos with a quiet merging
history after  $z\sim$2 can host disk galaxies (but see Springel $\&$
Hernquist 2005). However N-body simulations indicate that halos with a quiet
merging history since $z\sim$3 (those expected to host bulgeless disks), should
in general have too low an angular momentum to reproduce the observed
scale lengths of pure disks (D'Onghia $\&$ Burkert 2004).

The aim of this paper is to better understand  the {\it origin } of the
cosmological angular momentum problem. Our hope is that a more detailed
insight could help fo find a realistic solution for this unsolved 
puzzle.  The plan of this paper is as follows. In \S~\ref{sec:numexp} we present some
numerical simulations which are analysed in \S~\ref{sec:results} 
where we identify the dominant process that leads to the angular momentum
loss of the gas. \S~\ref{sec:conc} 
discusses the implications for our current understanding of the formation of
disk galaxies in a hierarchical universe.

\section{THE NUMERICAL METHOD}
\label{sec:numexp}

We perfomed a set of high--resolution simulations of a region that evolves to
form, at $z=0$, galaxy-sized dark matter halos in a low-density, flat,
``concordance'' Cold Dark Matter ($\Lambda$CDM) scenario: 
$\Omega_0=0.3$, $h=0.7$, $\Omega_{\rm b}=0.039$,
$\Omega_{\Lambda}=0.7$, $\sigma_8=0.9$. 
{\footnote{ We express the present value of Hubble's constant as
$H(z=0)=H_0=100\, h$ km s$^{-1}$ Mpc$^{-1}$}} 
The region was first identified in a low resolution cosmological simulation of a 
periodic box ($10 \, h^{-1}$ Mpc on a side) that was evolved until the present time
(z=0) using the Tree+SPH code GADGET2 (Springel 2005)  and then 
resimulated at higher resolution.
At $z=0$ the target dark matter halo under consideration has a circular velocity,
$V_{200}\sim 150$ km/s, and a total mass of $M_{200}=5 \times 10^{11} \, h^{-1} \,
M_{\odot}$, measured at the virial radius, $r_{200}=150 \, h^{-1}$ kpc, where
the mean inner density contrast (relative to the critical density for closure)
is $200$. 
This halo was selected from a list of clumps compiled using a
friend-of-friends algorithm with linking parameter  set to 15\% of the 
mean interparticle separation. 
The particles of the identified region were then traced back to the
initial conditions, where a box containing all of them was drawn.
The high-resolution region was filled with $256^3$ particles on a cubic grid
and the appropriate small-scale power was added up to the Nyquist frequency of
the new particle grid, preserving the Fourier amplitudes and phases of the low
resolution box. 
The outer regions were coarse-sampled using shells filled with particles of increasing
mass, up to the original resolution, in order to reproduce the original tidal field.
The re-simulation was performed using the multi--mass technique initial
condition implementation of the package ART (Kravtsov, Klypin \& Khokhlow
1997; Klypin et al. 2001).

A gas component was included by placing the same number of  gas particles on
top of the dark matter (DM) particles in the high-resolution box at the starting
redshift $z_i \sim 74$. The gas particles were then shifted by half the grid
size and their velocities were calculated by averaging the velocities of the 
eight neighbouring DM particles. ``Border'' grid cells (i.e. cells with less
than eight high-resolution DM particles) were left gas--empty.
For this simulation the number of high--resolution SPH+DM particles was $N_p \sim$
1.100.000.
The gas and dark matter particle mass was $m_{\rm g}=6.45\times10^5\, h^{-1}\,
M_{\odot}$ and $m_{\rm dm}=4.3\times 10^6\, h^{-1}\, M_{\odot}$,
respectively. This corresponds to a baryonic fraction $F_b=0.13$ consistent
with $\Omega_b=0.04$.
We adopted a comoving Plummer softening scale length of $1 h^{-1}$ kpc for all
gravitational interactions between pairs of particles. The minimum SPH
softening for the gas was fixed to half this value. 

We followed the strategy of keeping our model as simple as possible, hence
focusing our investigation on the effects of the presence of a collisional
component (gas) on the dynamical properties of the target halo, and in
particular on the evolution of the angular momentum in the collisional and in
the collisionless (DM) component. The affect of star formation and thermal 
feedback were neglected in this work and will be discussed in a subsequent paper. 
Instead, the
temperature of each gas particle was fixed at a
constant value of 10$^4$ K. Due to the high density of the cold gas 
at late times, all runs including gas were stopped at $z=0.3$, since the computational time
would have been too expensive. This choice does not affect the conclusions of this work
as all the important action which we plan to discuss happens at higher redshifts.

\section{RESULTS}
\label{sec:results}

\subsection{Properties of the target halos}
Previous work has explored the importance of mergers in explaining galactic
morphologies and has shown that major mergers with mass ratios 1:1
to 4:1, setup with cosmological self-consistent orbital parameters (Khochfar \& Burkert 2006),
 produce remnants with properties in agreement with observed
elliptical galaxies (Naab \& Burkert 2001; 2003; Burkert \& Naab 2004, 2005; Naab \& Trujillo 2005).
Disk galaxies are in contrast expected to form from mergers of substructures with
larger mass ratios (so called ``minor mergers'' and ``smooth accretion''). 
The more quiescent the merger history, the more likely it should be to form
extended disks. In addition, a high specific angular momentum would help.
We therefore identified dark halos in the low-resolution run
with a smooth merging history and high spin
in the original cube to be resimulated at higher resolution
with the following requirements: {\it i)} no
major merger with mass ratio $\leq 4:1$ from redshift $z=3$ until the present time. 
{\it ii)} high specific angular momentum of the halo as defined by the
dimensionless spin parameter $\lambda=J\sqrt{E}/GM^{5/3}$, where
$J$ is the total angular momentum, $E$ is the total energy, and $M$
is the total mass of the system.

Object A is the most quiescent halo we found in the original cube.
At z=0 it has a relatively large spin parameter
$\lambda\sim 0.04$, measured within the virial radius $\sim 150 \,h^{-1}$ kpc. 
We traced the merging activity of this halo backwards in time, following
the mass accretion history of the most massive progenitor as a function of
redshift. DM haloes at each simulation output were identified using a
Friends-of-Friends algorithm, with linking length $l=0.15$ times the mean
interparticle separation A major merger was assumed to occur if at
some time during $0 < z < 3$ the main progenitor was classified as a single
group in one output, but two separate groups with a mass ratio $\le 4:1$ in
the preceeding output (for details see D'Onghia $\&$ Burkert 2004). 

The redshift evolution of the mass within the virial radius
of the most massive progenitor of 
object A is shown in the upper panel of Figure 1.
Most of the merging activity of object A is over by z$\sim 1.5$.
A 1:1 merger actually occurs at z$\sim 3.8$. However
less than 15\% of the final mass is assembled by this time.
Most of the merging- and mass accretion activity happens at 
$z\sim 1.8$ when a 5:1 merger, coupled with substantial infall from 
filaments occurs. At that time the mass of
the system grows by 30\%, with only 20\% resulting from the 5:1 merger
and an additional 10\% being accreted simultaneously from the field.
After z$\sim 1.5$ the additional increase of
mass is due to minor mergers and accretion only.
Compared to the other halos, the mass accretion history of this
target halo is exceptionally quiet and should be
suitable for the formation of an extended, high-angular momentum disk.
For comparison, figure 1 also shows two additional halos, 
B and C,  with masses $10^{12} h^{-1}$M$_{\odot}$
and  $4$x$10^{11} h^{-1}$M$_{\odot}$ and high spin parameters of
$\lambda\sim 0.04$ and $\lambda\sim 0.03$ respectively.  Both
have also been selected in isolated regions of the original cube 
and resimulated with the same technique for comparison. 
Their merging activities however are more violent.
Halo B experiences a 2:1 major merger at
z$\sim$2.2. Halo C undergoes a 1:1 merger at z$\sim 2$.
For both halos, the merger
activity is largely over at z$\sim 1$ when most of the mass is in place.

Note that the selected halo A is a very rare exception due to its
large specific angular momentum of $\lambda \approx 0.04$, in
addition to its smooth merging history. 
If in this halo again a disk with small scale length would
form, the problem would be even worse in all other cases.
In a recent paper D'Onghia $\&$ Burkert (2004)
explored the angular momentum properties of halos that did not experience any
major mergers since z=3. They found that these quiescient halos
have low specific angular momenta (spin values)
that peak around a value of $\lambda\sim 0.02$, a factor of two lower
than the average in the log-normal distribution of spin values $\lambda =
0.04$ (Bullock et al. 2001)
and substantially smaller than expected for disk-dominated late-type galaxies
(van den Bosch, Burkert \& Swaters 2001)

\begin{figure}
 \psfig{figure=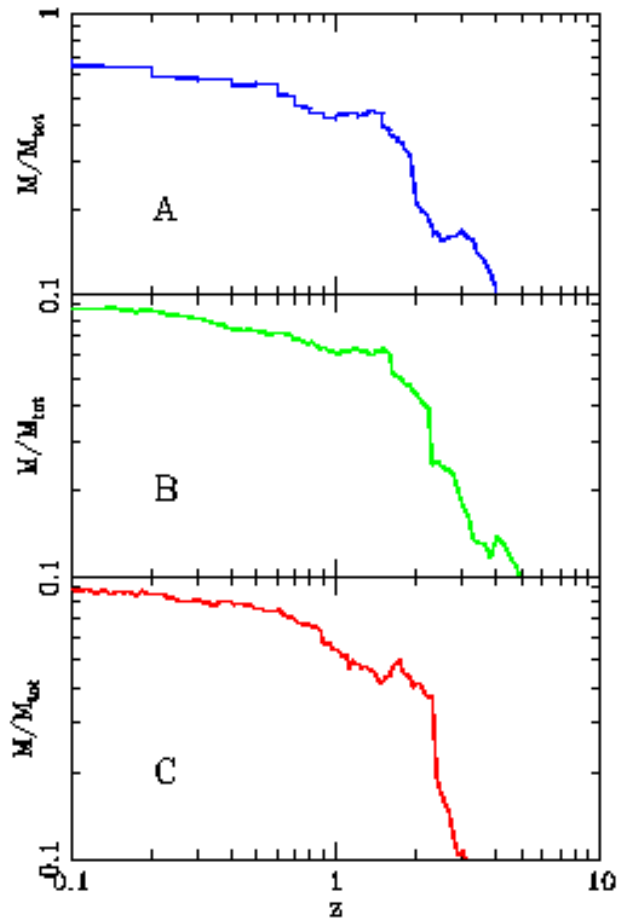,width=250pt}
  \caption{The mass fraction of the most massive progenitor, measured
  within the virial radius, with respect to
  the final mass at z$\sim$0, as a function of redshift for the target halos A, B
  and C.}\label{f1}
\end{figure}

\subsection{The origin of the compact disk component}

Does the gas retain its initial specific angular momentum during the evolution
of halo A without a major merger or is its angular momentum lost?

To answer this question let us focus on the evolution of object A.
Figure \ref{f2} shows the gas disk that has formed at $z=0.3$.
80\% of the gas particles within the virial radius are concentrated
in the central 3.5 $h^{-1}$ kpc, where they form a very compact disk of mass
$3.59\rm{x}10^{10} h^{-1}$M$_{\odot}$ (blue dots in Figure \ref{f2}) 
and scale length 1 $h^{-1}$  kpc, surrounded by a massive dark halo. 
The remaining 20\% of the gas particles have settled into a thin disk of
small total mass ($\sim 8.45\rm{x}10^{9} h^{-1}$M$_{\odot}$) (red dots
in Figure \ref{f2}) but scale length 2.5 $h^{-1}$  kpc ,
which extends to 10 $h^{-1}$ kpc from the center of mass of the remnant.
It is interesting that the outer lower-mass disk has a scale length that is
in good agreement with observations. The inner disk however is much
more compact than observed.
 
What is the origin of the compact disk? We find that during the 5:1 merger
at redshift 1.8 gas loses a substantial amount of its
specific angular momentum to the surrounding dark matter component due
to strong gravitational torques.
To quantify this process, we selected at z$\sim 0.3$ three groups
of particles: the dark matter particles within the virial radius; the gas particles 
in the inner disk within $\sim$3.5 $h^{-1}$ kpc from the galactic center
(blue dots in Figure \ref{f2}) and those between $3.5 < r < 10 \,h^{-1}$ kpc (red dots
in Figure \ref{f2} ). We traced back in time the evolution of the specific angular
momentum of each group with respect to its centre of mass.
Figure 3 (left panel), shows the specific angular momentum
for the dark matter (solid line), the inner gas particles (filled
circles) and outer gas particles (open circles)  as a function of
redshift. 
We find that the gas and the dark matter have similar angular momentum 
distributions until the 5:1 merger event at $z \sim 2$, when
a large fraction of the specific angular momentum of the gas, involved
in the merger is transferred to the surrounding dark matter.

Figure 4 shows the spatial distribution of all the material that at $z=0.3$
makes up the galaxy, traced back to z$\sim$2, just before the 5:1 merger:
the dark matter particles (black dots), the gas particles 
that are already in the disk of the major progenitor at $z=2$ (green dots) and the 
gas particles that are, at that time, in substructures or as diffuse
gas within the haloes of the progenitors  but that lateron fall into
the compact disk (blue dots). 50\% of the compact disk at z=0.3
(blue disk in Fig.\ref{f2}) 
is formed from gas particles that at z$\sim$2 are already in the 
disk of the more massive progenitor (green dots in figure 4).
13\% of the other 50\% of the gas (blue dots) is 
diffuse and in filaments, the rest is in clumps. During the merger event  
all that gas whether bound in substructures or diffuse 
but within the virial radius spirals inwards and loses angular momentum,
settling eventually into the compact disk.   

Why is the disk in the progenitor already that compact? Did it lose
specific angular momentum earlier?
The halo that contains the disk at z=2 assembled from a violent merging activity
at z$\sim$3.8 (see Fig. 1) when the density fluctuation corresponding to the progenitor 
collapsed.
To investigate the angular momentum distribution of this configuration we 
selected at z=2 the gas particles that form the disk of the major 
progenitor (green particles in figure 4) and its dark matter particles within
the virial radius and again followed back in time
their specific angular momenta. The right panel of  figure 3 
shows the evolution in redshift of the
specific angular momentum of these dark matter (blue
triangles) and gas particles (red triangles), respectively. Interestingly, between
$z=2$ and $z=5$, gas and dark matter have roughly the same, low specific angular
momentum. No signature of catastrophic angular momentum loss is found at
$z \sim 3.5$ which we attribute to the fact that due to their low angular momentum
the two progenitors in this 1:1 merger collide head-on
without generating gravitational torques that could remove angular momentum
if the objects would spiral in more gradually. We conclude that
the compact disk in the major progenitor
at $z \sim 2$ is not a result of angular momentum loss. Instead at the
time of its formation ($z \sim 3.5$) angular momentum build-up 
through gravitational interaction with neighboring structures
had not yet been efficient. The
angular momentum problem arises later, at $z=2$,
as a result of the violent merging
epoch when the system had gained a substantial amount of
angular momentum. Now the merger efficiently redistributes angular momentum
from the inner to the outer region.

\subsection{The origin of the extended disk component}

\begin{figure}
 \psfig{figure=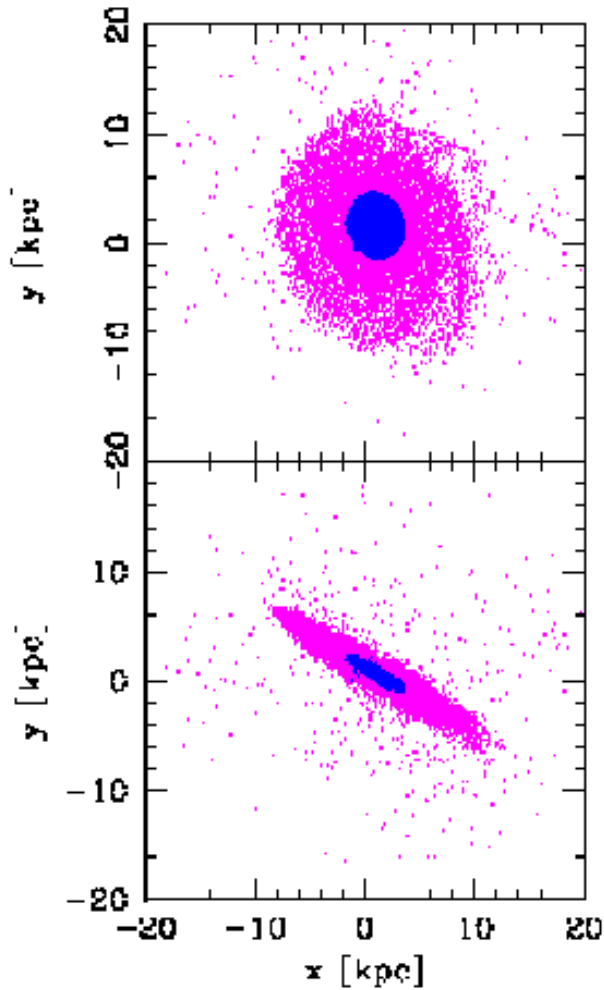,width=250pt}
  \caption{Gas particles projected such that the target object
A is viewed face-on (top panel) and edge-on (lower panel) at z$\sim$0.3.  
Almost 80\% of the gas particles 
within the virial radius  are concentrated
in the central disk of outer radius 3.5 $h^{-1}$ kpc (blue dots).
The remaining 20\% of the gas particles has settled into a thin, more
extended and lower mass disk (red dots).}\label{f2}
\end{figure}

What is the origin of the extended thin disk at z=0.3 in Figure \ref{f2} ?
It formed after the 5:1 merger
event, by smooth accretion of gas ($\sim$ 30\%)and by gas stripped 
from infalling dark matter substructures with mass ratios larger 
than 10:1 ($\sim$ 70\%).
As demonstrated in Figure 4 this
gas (red dots) is still too far away at z=1.8 to lose angular momentum
by the gravitational torques that arise in the course of the 5:1 merger.
In fact, like the extended dark matter component, it gains angular momentum
transferred from the inner, merging parts to the outer regions.
This is shown by the red open circles in Figure 3 (left panel). 

\begin{figure}
 \psfig{figure=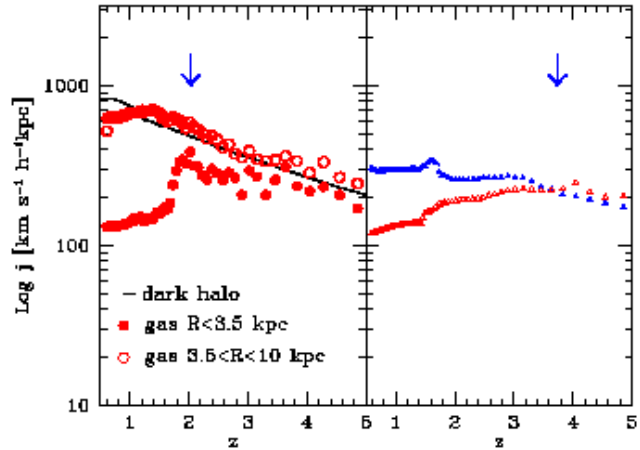,width=250pt}
  \caption{The left panel shows the specific angular momentum of
the dark matter within the virial radius (solid line), of the gas
particles  that are placed within 3.5 $h^{-1}$ kpc from the center of
mass at z$\sim$0.3 (filled
circles) and of the gas particles that lie at radii
$3.5 < r < 10 \,h^{-1}$ kpc at z$\sim$0
(open circles)  as a function of redshift. 
The curves
in the right panel show the angular momentum evolution 
of the dark matter (blue) and the gas (red) of the massive
progenitor that at z=2 has already formed a compact disk.}\label{f4}
\end{figure}

\begin{figure}
\psfig{figure=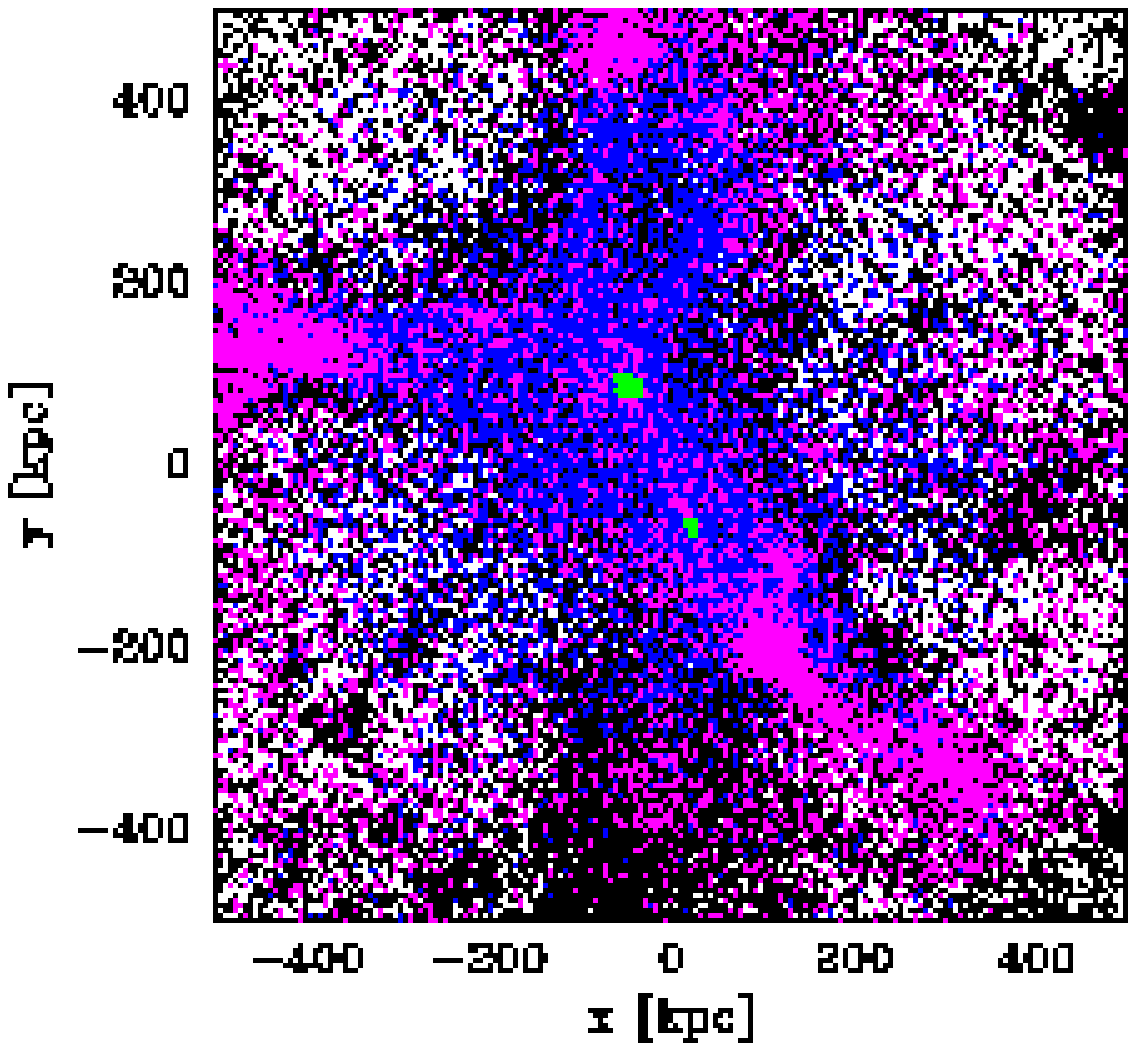,width=250pt}
\caption{The spatial distribution of all the material that at
z=0.3 forms the galaxy traced back to z$\sim$2: 
the dark matter (black dots), the
gas particles that at z=0.3 form the inner disk (blue dots) and the
gas particles that at z=0.3 form the outer extended disk
(red dots). The green dots show the dense, small disk
of gas that has already formed in the center of the
more massive merger components at z=2.}\label{z2}
\end{figure}

\begin{figure}
 \psfig{figure=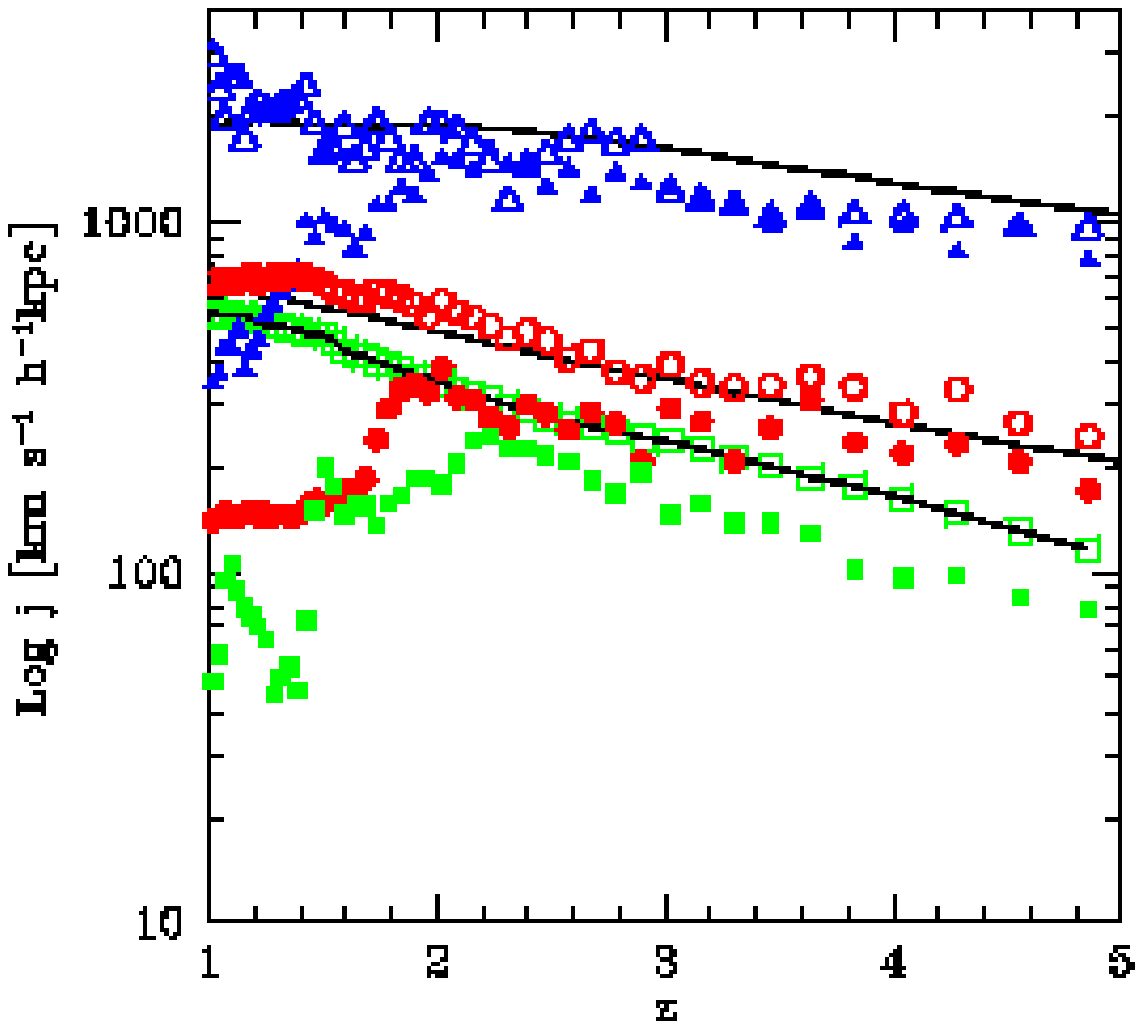,width=250pt}
  \caption{The evolution in redshift of the specific angular momentum
of the dark matter within the virial radius (solid lines), the cold
gas in the compact disk (filled symbols) and the gas in
the extended disk (open symbols) is shown for the halos
A (middle), B (top)
and C (bottom).}\label{f6}
\end{figure}

Similarly to halo A, the objects B and C also form 
a very compact disk of gas and subsequently accrete an extended low mass
disk of gas around it.
In Figure \ref{f6} the time evolution of specific angular momentum
of the gas and the dark matter for object B 
(blue symbols in the top panel) 
and object C (green symbols, bottom panel)  is compared with the
quiescent halo  A (red symbols, middle panel).  Model B,
during its 2:1 major merger at $z \sim 2.2$ assembles 50\% of its final mass and 80\% of gas loses
its angular momentum in this process. The same happens
to model C at $z \sim 2$ when a 1:1 merger occurs.
The filled triangles (object B) and filled squares (model C)
in Figure \ref{f6} show the time evolution of the
specific angular momentum of the gas that forms the
inner compact disk. The open symbols show the specific angular momenta of the
extended disks (20\% of total gas mass) that are accreted and gain
angular momentum.

\section{Conclusion}
\label{sec:conc}
The origin of the angular momentum loss of 
gas in cosmological simulations of galaxy formation has been investigated.
We focussed on the most promising candidate for the formation of
an extended disk component: a halo with high spin and
no major merger since redshift 3. The gas and the dark matter initially
have similar angular momenta. However at $z \sim 2$, the gas
involved in a 5:1 merger loses most of its angular momentum 
by gravitational torques. Angular momentum loss during major
mergers with mass ratios of 1:1 to 4:1 is a
process that is well known and that has been studied in
simulations of elliptical galaxy formation 
(Mihos \& Hernquist 1996, Naab \& Burkert 2001). 
Interestingly however, even less violent (substantial) mergers
with mass ratios of order 4:1 to 6:1 that are not
considered to produce elliptical galaxies appear to be still very
efficient in redistributing specific angular momentum 
(Bournaud et al. 2005). Semi-analytic models predict these mergers
to be an order of magnitude more frequent than major mergers 
(Khochfar \& Silk 2005).

Contrary to common believe, dynamical friction that would always act
during clumpy gas infall is not the dominant cause of the angular momentum problem.
For example, the gas that falls in after the last substantial merger, although
it resides in dark matter substructures, does not lose its 
specific angular momentum, contrary to what one would
expect if dynamical friction is the reason for angular momentum loss.
Instead it is the 5:1 merger
that occurs when a large fraction of the galaxy has been
assembled that generates the angular momentum problem.
We also checked that the dark matter in the inner region
experiences a similar angular momentum loss as the gas which confirms
that gravitational torques are the dominant process. 

In the simulations
presented here, there was no dark halo which had an even more
quiet merging history. In
a subsequent paper we will present a detailed statistical investigation
of galaxy merging histories, based on 
semi-analytical models by Khochfar \& Burkert (2001, 2005)
that confirms the present conclusion and demonstrates
that the likelihood for a disk galaxy to {\it not} experience at
least one substantial 
merger event since redshift 3 is extremely small. Rather than being
a dynamical friction issue, the angular
momentum problem therefore appears to arise from the fact
that in the cold dark matter scenario substantial mergers are expected
to occur in almost all galaxies.

An interesting scenario of disk galaxy formation has been proposed by 
Birnboim \& Dekel (2003) and  Keres et al (2005) who suggested
that gas in high redshift disk galaxies is accreted cold and without  
virial shocks directly from filaments, forming galactic disks.
Our simulations demonstrate however that this 
scenario will also suffer from the same angular problem as other models,
as a substantial merger is likely to destroy these early disks lateron.

We find that the gas outside the merging region that is
accreted lateron keeps most of its
angular momentum, forming an extended secondary disk with a scale
length that is in excellent agreement with observations.
Typical galactic disks therefore can form naturally in 
galaxy formation simulations. However, these disks should form late
and should only contain a small baryon fraction.
Most of the baryons should
reside in the center where they form large bulges.
This is consistent with the cosmological simulations e.g. of Abadi et al. 2003.

What is then the origin of the late-type spiral galaxy population
with small bulge-to-disk ratios? We can think of several possible
answers.
First, most of the gas that settles into the inner disk might
be blown out by a central supermassive black hole (Robertson et al. 2005, astro-ph/0503369),
leaving behind a small bulge that however
needs to be large enough to fit the observed black hole mass versus bulge mass
correlation (H\"aring  et al. 05). This bulge would be surrounded by
the now dominant, extended disk component that formed from late infall.
Note that this scenario could explain why spiral galaxies have
baryon fractions that appear to be smaller than the cosmologically
predicted universal baryon fraction (Yang et al. 2005).
Unfortunately, the problem of generating extended disks in bulgeless galaxies cannot be
solved in this way (D'Onghia $\&$ Burkert 2004).

Another process that could reduce the mass of the inner disk
is energetic feedback that decouples the
gas from the dark matter and prevents it from falling in
before the last substantial merger episode. This might explain
why some numerical simulations with feedback are successful in 
generating dominant large disks while others fail. If gas heating occured
just before the last substantial merger, an extended disk
could form through subsequent monolithic gas infall.
If gas heating  would however act too early,
the gas would cool again and fall back into the inner regions 
before the merger occurs, leading again to an angular momentum problem.
Vice versa, if heating would be turned on too late, gas would already
have lost its angular momentum in the merger. This suggests that
the energetic feedback should best be coupled directly to
the last merger event e.g. by triggered star bursts and/or nuclear
activities, coupled with central black hole formation.

Whether one of these scenarios will solve the
angular momentum problem and can reproduce the observed frequency and
luminosity function of late type galaxies is an interesting question 
that should be explored in greater details.

\section*{Acknowledgements}
We thank Volker Springel for making GADGET2 available prior to publication
and Anatoly Klypin for  kindly providing the ART IC package.
We also thank Julio Navarro for many interesting discussions.
SK acknowledges funding by the PPARC Theoretical Cosmology Rolling
Grant.

\label{lastpage}


\begin{thebibliography}{}

\bibitem[Abadi et al. (2003)]{a03}
  Abadi~M., Navarro~J.~F., Steinmetz~M., Eke~V.~R. 2003, ApJ, 591, 499

\bibitem[Birnboim \& Dekel (2003)]{BD03}
  Birnboim, Y., Dekel, A. 2003, MNRAS, 345, 349

\bibitem[Blumenthal et al. (1984)]{B84}
Blumenthal~G.~R., Faber~S.~M., Primack~J.~R., Rees~M.~J. 1984, Nature, 311, 517

\bibitem[Bournaud et al.(2005) Bournaud et al.]{bj05}
Bournaud~F., Jog~C.~J. \& Combes~F. 2005, AA, 437, 69

\bibitem[Bullock et al.(2001) Bullock et al.]{b01}
{Bullock~J.~S., Dekel~A., Kolatt~T.~S., Kravtsov~A.~V., Klypin~A.~A., Porciani~C., Primack~J.~R. 2001, ApJ, 555, 240}

\bibitem[Burkert \& D'Onghia (2004) Burkert]{burkert04}
{Burkert~A. \& D'Onghia, E. 2004, in "Penetrating Bars Through
Masks of Cosmic Dust", eds. D. Block, E. Block, K. Freeman, K. Puerari, R. Groess (Springer), p. 341, astro-ph/0409540}


\bibitem[Burkert \& Naab(2004)]{BN04}
{Burkert~A., Naab~T. 2004, in ``Coevolution of Black Holes and Galaxies'', 
eds. L. C. Ho, p. 421} 

\bibitem[Burkert \& Naab(2005)]{BN05}
{Burkert~A., Naab~T. 2005, MNRAS, 363, 597} 

\bibitem[Courteau(1997)]{C97}
{Courteau~S. 1997, AJ, 114, 240}

\bibitem[D'Onghia \& Burkert (2004) D'Onghia \& Burkert]{db04}
{D'Onghia~E. \&  Burkert~A. 2004, ApJ, 612, L13}





\bibitem[Fall \& Efstathiou(1980) Fall \& Efstathiou]{fe80}
{Fall~S.~M. \& Efstathiou~G. 1980, MNRAS, 193, 189}

\bibitem[Governato et al.(2004) Governato et al.]{G04}
{Governato~F., Mayer~L., Wadsley~J., Gardner~J.~P., Willman~B., Hayashi~E., Quinn~T.,
 Stadel~J., Lake~G. 2004, ApJ, 607, 688}

\bibitem[H\"aring \& Rix (2004)]{HR04}
{H\"aring~N. \& Rix~H-W. 2004, ApJ, 604, L89}


\bibitem[Keres et al(2005)]{K05}
{Keres~D., Katz~N., Weinberg~D.H., Dave~R. 2005, MNRAS, 363, 2}

\bibitem[Khochfar \& Burkert(2001)]{2001ApJ...561..517K} 
  Khochfar~S., \& Burkert~A.\ 2001, ApJ, 561, 517 

\bibitem[Khochfar \& Burkert(2003)]{2003ApJ...597L.117K} Khochfar, S., \& 
Burkert, A.\ 2003, ApJ, 597, L117 
 
\bibitem[Khochfar \& Burkert(2005)]{2005MNRAS.359.1379K} Khochfar, S., \& 
Burkert, A.\ 2005, MNRAS, 359, 1379 

\bibitem[Khochfar \& Silk(2005)]{ks05} Khochfar, S., \& 
Silk, J.\ 2005, ArXiv Astrophysics e-prints, arXiv:astro-ph/0509375 

\bibitem[Khochfar \& Burkert(2006)]{KB06}
 Khochfar~S. \& Burkert~A. 2006, A\&A, 445, 403 

\bibitem[Klypin et al.(2001)]{K01}
{Klypin~A., Kravtsov~A.~V., Bullock~J.~S., Primack~J.~R. 2001, ApJ, 554, 903}

\bibitem[Kravtsov et al.(1997)]{K97}
{Kravtsov~A.~V., Klypin~A.~A., Khokhlov~Alexei~M. 1997, ApJS, 111, 73}



\bibitem[Mihos \& Hernquist]{MH96}
{Mihos, J.C, Hernquist, L. 1996, ApJ, 464, 641}

\bibitem[Mo, Mao \& White(1998) Mo, Mao \& White]{mo98}
{Mo~H.~J., Mao~S. \& White~S.~D.~M. 1998, MNRAS, 295, 319}

\bibitem[Naab \& Burkert(2001)]{NB01}
{Naab~T., Burkert~A 2001, ApJ, 555, L9}

\bibitem[Naab \& Burkert(2003)]{NB03}
{Naab~T., Burkert~A 2003, ApJ, 597, 893}

\bibitem[Naab \& Trujillo(2005)]{2005astro.ph..8362N} Naab, T., \& 
Trujillo, I.\ 2005, ArXiv Astrophysics e-prints, arXiv:astro-ph/0508362 

\bibitem[Navarro \& Benz (1991) Navarro \& Benz]{n91}
{Navarro~J.~F., Benz~W. 1991, ApJ, 380, 320}




\bibitem[Navarro, Frenk \& White(1995) Navarro, Frenk \& White]{nfw95}
{Navarro~J.~F., Frenk~C.~S. \& White~S.~D.~M. 1995, MNRAS, 275, 720}




\bibitem[Navarro \& Steinmetz (1997) Navarro \& Steinmetz]{n97}
{Navarro~J.~F., Steinmetz~M. 1997, ApJ, 478, 13}


\bibitem[Navarro \& Steinmetz (2000) Navarro \& Steinmetz]{n00}
{Navarro~J.~F., Steinmetz~M. 2000, ApJ, 538, 477}

\bibitem[Peebles (1969) Peebles]{P69}
{Peebles~P.~J.~E. 1969, ApJ, 155, 393}
 
\bibitem[Primack (2005) Primack]{p05}
{Primack, J.R. 2005, NewAR, 49, 25}

\bibitem[Robertson et al.(2004) Robertson et al.]{r04}
{Robertson~B., Yoshida~N., Springel~V., Hernquist~L. 2004, ApJ, 606, 32}

\bibitem[Robertson et al.(2005)]{R05}
{Robertson~B., Hernquist~L., Bullock~J.S., Cox~T.J., Di Matteo~T., 
Springel~V., Yoshida~N. 2005, ApJL submitted, astro-ph/0503369} 


\bibitem[Sommer-Larsen et al.(2003) Sommer-Larsen et al.]{sl03}
{Sommer-Larsen~J., Gotz~M., Portinari~L. 2003, ApJ, 596, 47}

\bibitem[Springel (2005) Springel]{Sp05}
{Springel~V. 2005, MNRAS, 999}

\bibitem[Springel \& Hernquist]{SH05}
{Springel~V. \&  Hernquist~L. 2005, ApJ, 622, L9}


\bibitem[Steinmetz \& Mueller(1994)]{SM94}
{Steinmetz~M. \& Mueller~E. 1994, A\&A, 281, L97}

\bibitem[van den Bosch et al(2001)]{vB01}
{van den Bosch, F.C., Burkert, A., Swaters, R.A. 2001, MNRAS, 326, 1205}

\bibitem[White \& Rees(1978)]{W78}
{White~S.D.M. \& Rees~M.~J 1978, MNRAS, 183, 341}

\bibitem[White (1984) White]{w84}
{White~S.~D.~M 1984, ApJ, 286, 38}

\bibitem[Yang et al.(2005)]{Y05}
{Yang~X., Mo~H.J., Jing~Y.P., van den Bosch~F.C. 2005, MNRAS, 358, 217}

\end{thebibliography}
\end{document}